# The Black Hole Evolution and Space Time (*BEST*) Observatory

Response to the Request for Information "Concepts for the
Next NASA X-ray Astronomy Mission" (NNH11ZDA018L)

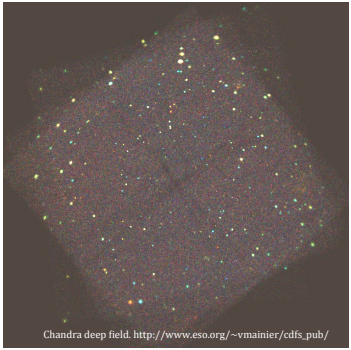 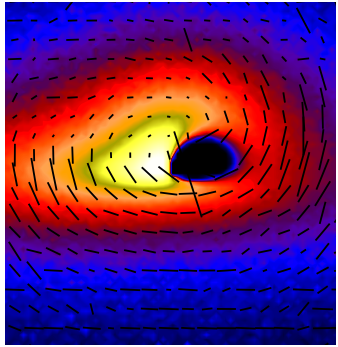 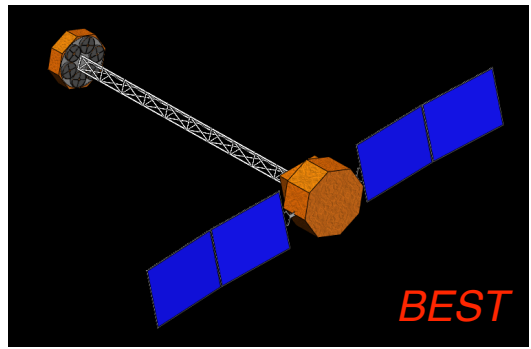

*Response Category:* Mission Concept - *BEST* is a mission for deep hard X-ray imaging (5-70 keV) and broadband X-ray polarimetry (2-70 keV) for 582 M$ total costs including contingency.

*Lead:* Professor Henric Krawczynski, Washington University in St. Louis,

| Team Members | Affiliation |
|---|---|
| Jack Tueller | GSFC |
| Scott Barthelmy | GSFC |
| Jeremy Schnittman | GSFC |
| William Zhang | GSFC |
| Julian Krolik | Johns Hopkins Univ. |
| Matthew G. Baring | Rice University |
| Ezequiel Treister | Univ. of Hawaii |
| Richard Mushotzky | Univ. of Maryland |
| Matthias Beilicke | Washington Univ. |
| James Buckley | Washington Univ. |
| Ram Cowsik | Washington Univ. |
| Martin Israel | Washington Univ. |

## *BEST* Science Matrix

| Scientific Goal | *BEST* Measurements | Instrument Requirements |
|---|---|---|
| 1.) Study the behavior of matter and spacetime close to the event horizons of black holes. | Time and energy resolved polarimetry observations of galactic black holes, AGNs, and AGN jets. | Broadband polarimetry 2-70 keV with <1% MDP for galactic and extragalactic sources. |
| 2.) Measure the evolution of supermassive black holes from $z=0$ to $z>6$. | AGN hard X-ray survey down to $F(6-10\text{ keV})=10^{-16}$ erg cm$^{-2}$ s$^{-1}$. | Field of view 12′ (FWHM); 6-10 keV sens. = $10^{-16}$ cgs; ang. res. (HPD) ≤ 10″. |
| 3.) Map large scale structure. | AGN survey, and maps of the non-thermal emission from large scale structure shocks. | Field of view 12′ (FWHM). At 6 keV: 3000 cm$^2$ mirror & $<2\times10^{-3}$ cm$^{-2}$ s$^{-1}$ keV$^{-1}$ background. |
| 4.) Study how AGNs affect their hosts. | Study AGN bubbles in galaxy clusters, and AGN-galaxy co-evolution. | Requirements of 2) and 3). |
| 5.) Constrain the equation of state of neutron stars and study their magnetospheres. | Hard X-ray polarimetry observations of neutron stars and magnetars. | Same as 1. |

## *BEST* Mission Concept

| Mission Component | Requirements | Technology of Choice/ Comments |
|---|---|---|
| X-ray telescope and mirror assembly | Mirror with 3000 cm$^2$ area at 6 keV, 2-70 keV band-pass, angular res. ≤ 10″. | 6 mirror ass., each: Ø50 cm, 600 cm$^2$ (2 keV), 500 cm$^2$ (6 keV), 250 cm$^2$ (20 keV). |
| Hard X-ray imaging detector | 5-70 keV, cover 3.5×3.5cm$^2$ area, 120-240 μm pixels, ΔE/E = 1.5 keV FWHM. | 2 mm thick CZT cross strip detectors. |
| Broadband X-ray polarimeter | 2-70 keV, high detection efficiency. | Photoelectric-effect+ Compton-effect broadband polarimeter. |
| Spacecraft & launch vehicle | Continuous roll to reduce the systematic errors of the polarization measurements. | Taurus-class launcher and spinning spacecraft. |
| Total mass, power, cost | 1040 kg, 581 W, 582 M$. | 3 (5) year base mission. |

## 1. Executive Summary

*The BEST (Black hole Evolution and Space Time) mission will make use of a 3000 cm² effective area mirror (at 6 keV) to achieve unprecedented sensitivities for hard X-ray imaging spectrometry (5-70 keV) and for broadband X-ray polarimetry (2-70 keV). BEST will make substantial contributions to our understanding of the inner workings of accreting black holes, our knowledge about the fabric of extremely curved spacetime, and the evolution of supermassive black holes.*

Similar to *IXO, BEST* will allow for time resolved studies of accretion disks. With a 7.6 times larger mirror area and a 7 times wider bandpass than *GEMS*[1], *BEST* will take X-ray polarimetry to a new level: it will probe the **time variability of the X-ray polarization from stellar mass and supermassive black holes**, and it will measure the polarization properties in ~30 independent energy bins (compared to *GEMS*'s 2-4 independent energy bins). These capabilities will allow *BEST* to conduct precision tests of accretion disk models and the underlying spacetime. With 3 times larger mirror area and 10 times better angular resolution than *NuSTAR*[2], *BEST* will be able to make deep field observations with a >15 times better sensitivity than *NuSTAR*. The mission will be able to trace the **evolution of obscured and unobscured black holes in the redshift range from z=0 to z>6**, covering the most important epoch of supermassive black hole growth. The hard X-ray sensitivity of *BEST* will enable a deep **census of non-thermal particle populations**. *BEST* will give us insights into AGN feedback by measuring the particle luminosity injected by AGNs into the interstellar medium (ISM) of their hosts, and will map the emission from particles accelerated at large scale structure shocks. Finally, *BEST* has the potential to constrain the **equation of state of neutron stars** (NS).

*Table 1:* BEST Performance.

| Mirror | Requirement (Goal) |
|---|---|
| Area | 3,600 cm² at 2 keV, 3,000 cm² at 6 keV, 1,500 cm² at 20 keV. |
| Angular Resolution | 10″ (5″) Half Power Diam. |
| Field of View (FWHM) | 12′ (20′) |
| **Imager** | |
| Bandpass | 5-70 (5-90) keV |
| Sensitivity (10⁶ s) | 1 × 10⁻¹⁶ cgs (6-10 keV), 3 × 10⁻¹⁶ cgs (10-20 keV), 2 × 10⁻¹⁵ cgs (20 -70 keV) |
| Energy Resolution | 1.5 (1) keV FWHM |
| **Polarimeter** | |
| Bandpass | 2-70 (2-90) keV |
| Sensitivity (10⁶ s, 1 mCrab source) | 7‰ MDP (2-10 keV) 3% MDP (10-70 keV) |
| Energy Resolution | 20% below 10 keV, ~2 keV above 10 keV |

*BEST leverages breakthroughs in lightweight high-precision mirror technology developed for IXO and novel focal plane technologies to achieve a performance vastly superior to NuSTAR and GEMS.* The mission will use six mirror assemblies (similar to the two mirror assemblies of *NuSTAR*) on a deployable boom. The six mirrors will focus the X-rays either on six imaging spectrometers or on six polarimeters. For a three year mission we estimate a total mission cost of 582 M$ (incl. contingencies).

## 2. The Science Case of BEST

*BEST* will contribute substantially to answering the *IXO* science questions. The description of the science potential is based on the performance specifications listed in Table 1. The sensitivity of the hard X-ray imager was estimated based on our predictions of the mirror area and the mirror point spread function, and on the background level from *NuSTAR* simulations[2]. The minimum detectable polarization (MDP) of the broadband polarimeter was estimated based on scaling the *GEMS* sensitivity with mirror area and on detailed simulations of the *X-Calibur* experiment[3]. In the following we discuss the *BEST* contributions to the science questions.

**2.1 What happens close to a black hole?** *BEST* will allow us to use accreting black holes (BHs) as precision laboratories for testing accretion disk models and probing the spacetime structure around BHs.

Even though Einstein's theory of general relativity (GR) has passed high-precision tests in the weak-gravity regime of solar system and binary pulsar observations, its predictions in very deep gravitational potentials (large $\varepsilon = GM/rc^2$ ) and strongly curved space-times (large $\xi = GM/r^3c^2$) have not yet been tested adequately[4]. To take full advantage of X-ray observations of accreting BHs and test GR at large $\xi$ and $\varepsilon$, a detailed understanding of accretion disk physics is needed. Recently, some progress has been made in comparing GR magneto-hydrodynamic simulations with analytic models and observations of X-ray spectra, yet many questions remain unanswered[5, 6]. X-ray polarimetry will directly probe the geometry of the accretion flow around BHs (Fig. 1) and can be used to break degeneracies in the models that limit the effectiveness of purely spectral observations[7, 8]. With a collecting area 7.6 times larger than *GEMS*, *BEST* will be able to carry out **dynamic** polarization measurements of bright sources, phase-resolving the light curves to understand the physical mechanisms behind the mysterious quasi-periodic oscillations(QPOs) seen in many galactic BHs[9]. The large collecting area will allow *BEST* to reach much greater sensitivity than *GEMS* for AGNs with typical X-ray fluxes in the 1-10 mCrab range, thereby increasing both the number and diversity of observable sources. For both AGN and galactic sources, the degree of polarization is expected to increase with energy[7, 8], so that *BEST*'s broader bandpass will be another significant advantage over *GEMS*. *BEST* 2-70 keV polarimetry of AGN will test models of the structure of magnetized coronae and jets, with implications for how AGN evolve through a wide range of accretion processes[10].

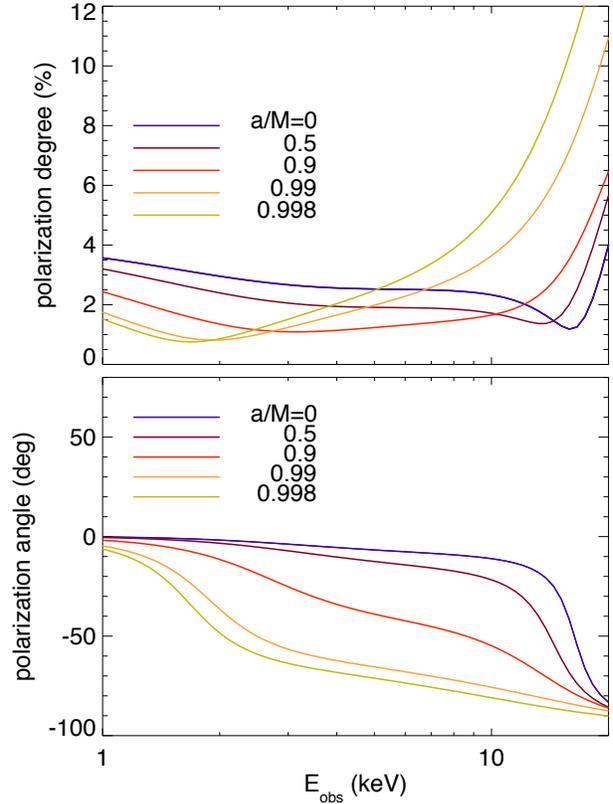

*Figure 1:* Models of the polarization degree and direction of the thermal-state-emission from an accreting stellar mass black hole for several values of the dimensionless spin a/M (angular momentum per unit mass), demonstrating the power of X-ray polarimetry to measure the black hole spin and to test the underlying model assumptions[7, 8]. For bright sources, the *BEST* errors will be much smaller than the differences between the models.

**2.2 When and how did supermassive black holes grow?** With a sensitivity >15 times better than *NuSTAR*, a <10″ angular resolution, and hard X-ray capabilities, *BEST* can observe even the most heavily obscured black holes and determine their evolutionary history in the *z*=0 to *z*>6 redshift range, detecting ~380 sources in a $10^6$ s pointing (Fig. 2).

A large fraction of the accreting supermassive black holes are heavily obscured and will be inaccessible to detailed observations in the optical, UV and soft X-ray bands. A recent stacking analysis already shows that the *Chandra* deep field observations are missing large numbers of heavily obscured Compton-thick AGNs[11]. Hard X-ray emission is a generic feature of accreting black holes and the intensity is *directly proportional* to the bolometric luminosity; thus *BEST* observations will improve over the luminosity estimates based on IR data, which are sensitive to the emitted spectrum and the geometry and composition of the matter surrounding the BH[12]. The broad spectral bandpass of *BEST* will make it possible to measure the column-density and other properties of the obscuring material even for extremely obscured sources. *BEST*'s angular resolution will enable unambiguous identification and will avoid confusion.

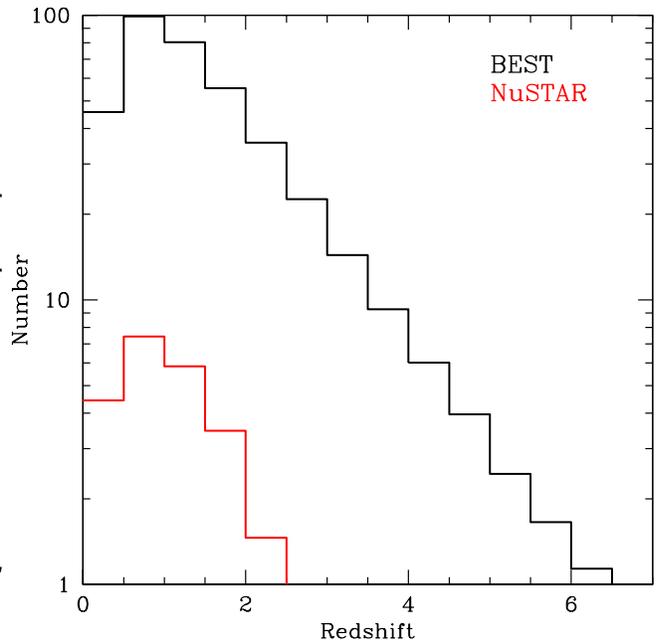

*Figure 2:* Predicted AGN detections with *NuSTAR* (lower red curve) and *BEST* (upper black curve) for a $10^6$ s deep field observations. The *BEST* survey will detect the obscured objects missed by *Chandra* and will be substantially deeper than the *NuSTAR* survey (see[11, 12]).

### 2.3 How does large scale structure evolve and how is it connected to supermassive black hole formation?

*BEST* observations will give us information about large scale structure, and the co-evolution of galaxy clusters, galaxies and black holes. *BEST* will have unprecedented sensitivity for inverse Compton emission from non-thermal electrons making it possible to detect electrons accelerated (i) at large scale structure formation shocks (and thus to map large scale structure) and (ii) in galaxy cluster bubbles (with implications for how AGNs heat the cluster gas). The *BEST* AGN census will cover obscured and unobscured AGNs and will enable a study of the co-evolution of AGNs and their hosts.

### 2.4 How does matter behave at very high density? *BEST* will help to constrain the equation of state (EOS) of neutron stars by measuring the spectra and polarization of radiation from their surfaces and atmospheres.

Clean measurements of the surface thermal emission are needed to characterize the temperature and to constrain both their *masses* and *radii*. X-ray outbursts from low-mass X-ray binaries (XRBs) are prime candidates for such studies[13, 14], their temperatures being at the low end of the *BEST* window. The NS radius will be calibrated using multiple flux measurements combined with the Stefan-Boltzmann law. Some outbursts exhibit signatures of photospheric radius expansion (PRE), which is bounded by a surface "touchdown" Eddington-limited flux that constrains the mass-to-radius ratio. The combined mass/radius diagnostics probe the EOS. *BEST* will be able to refine these determinations for select X-ray bursters using polarization information to tighten the emission locale constraints. Moreover, because of its hard X-ray sensitivity, *BEST* will be able to extend such analyses to the domain of magnetars that exhibit outbursts above 10 keV. Prospects for the identification of PRE in magnetar bursts in the hard X-ray band are good[15], and constraints on the EOS in magnetars (a first) would provide key insights into their relationship with XRBs and pulsars.

## 3. Technical Description

**3.1 Overview:** We envision a baseline mission of 3-5 years in a 600 km circular orbit (28° inclination). The scientific payload consists of two components: six mirror collectors on a deployable boom of 10 m length, and the imaging and polarimetry focal plane instrumentation on a turntable (Fig. 3).

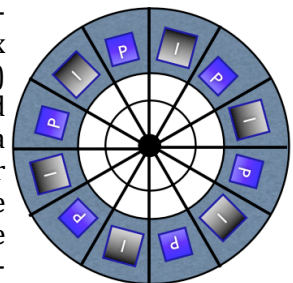

*Figure 3:* The *BEST* focal plane instrumentation consists of six hard X-ray imagers (I) and six broadband polarimeters (P) on a turntable with an outer diameter of 70 cm. The table moves either the imagers or the polarimeters into focus.

**3.2 Imaging detectors:** *BEST* will use segmented CdZnTe detectors to minimize the cost of the hard X-ray focal plane instrumentation. In order to properly sample the mirror PSF, the pixel size of the focal plane detector must be approximately half of the half power diameter (HPD). A PSF with a HPD of 10″ (5″) should thus be combined with a pixel spacing of 240 μm (120 μm). Currently, the best pixel detectors with good energy resolution have pixel sizes of 600 μm[16], but progress in the pixel readouts is expected. We have shown that a spatial resolution of <100 μm can be achieved with CdZnTe 2-D strip detectors[17, 18] (Fig. 4). To cover ≥12′ FoV, the focal plane must be ≥3.5 cm on a side (slightly larger than *NuSTAR*), achievable with a mosaic focal plane of 4 parts. For a 12′ FoV, the detector mosaic requires 292 (584) ASIC readout channels. Active anti-coincidence shielding (CsI or BGO) and depth-of-interaction detection will be used to suppress background (as in *NuSTAR* and *InFOCµS*). The main effort will be the calibration and flat-fielding of the CdZnTe detectors. The imagers do not need new technology.

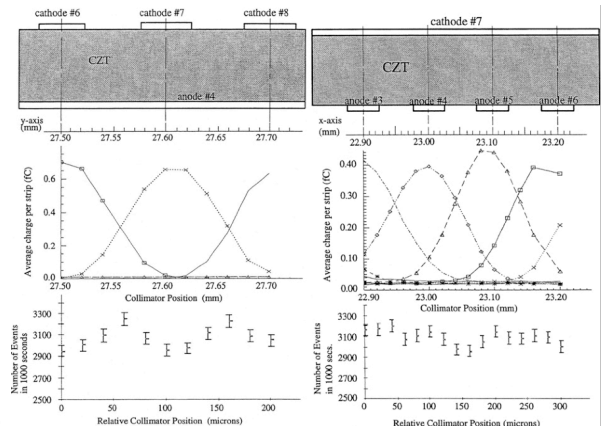

*Figure 4:* Linear scans of a 100 μm strip detector with a 30 μm collimated spot (59.5 keV line of $^{241}$Am). The center panels show the average signal as a function of spot position. The bottom panel shows the average count rate. The small variations in rate reflect a charge spreading that is >1 strip width not variations in efficiency.

**3.3 Broadband X-ray Polarimeter:** Each of the six X-ray polarimeters consists of two sections (Fig. 5): a "front" low-energy photoelectric-effect polarimeter (2-10 keV) followed by a "rear" high-energy Compton-effect polarimeter (10-70 keV). The polarimeters have high detection efficiency but no imaging capabilities. The design of the **low-energy polarimeter** will be almost identical to the ones used on the *GEMS* mission. Each unit will consist of four time projection chambers (TPCs) each with an active depth of 7.8 cm along the X-ray beam and a 3×3 cm$^2$ active cross section perpendicular to it, 128 readout strips, and a gas electron multiplier (Fig. 6)[1]. The amplified signals are sampled by one 128-channel APV25-ASICs with a sampling frequency of 20 MHz. The cathode signal is used to trigger the readout, to measure the energy of the detected X-rays, and to reject background events. Following a trigger, the ASICs deliver 30 samples per channel. Filled with 0.25 atm dimethyl ether (DME) the 4 TPCs absorb 99% of 2 keV photons and 10% of 8 keV photons. The **high-energy polarimeter** adopts the design of the *X-Calibur* experiment[3,19], a hard X-ray polarimeter scheduled for a 1-day balloon flight from Fort Sumner (NM) in 2014. The polarimeter uses a 14 cm long tapered plastic scintillator rod (read out by a PMT) to detect Compton interactions. The scintillator trigger can be used to reduce background at the expense of raising the energy threshold. The Compton-scattered photons are photo-absorbed in a "cage" of 32 Cadmium Zinc Telluride (CZT) detectors surrounding the scattering rod at the four long sides (Fig. 7). Each CZT detector (0.2×2×2 cm$^3$) has 64 pixels and will be read out with one 128-channel ASIC which fits below the footprint of the detector. The ASIC measures for each pixel the signal amplitude and the time offset between the pixel trigger and a common cathode trigger. The latter can be used to estimate the depth of interaction, and to suppress background.

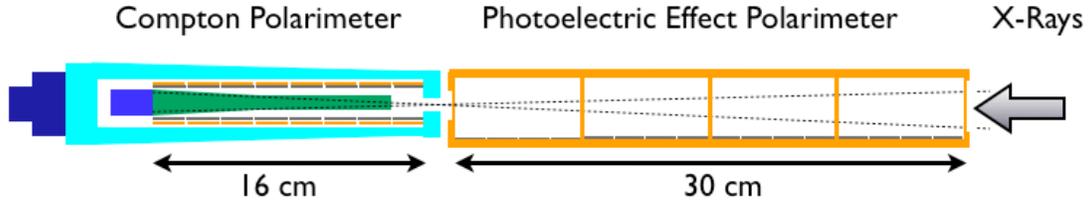

*Figure 5:* *BEST* will use a combination of photoelectric effect polarimeter and Compton polarimeter to cover the energy band from 2 keV to 70 keV. The X-rays enter from the right side of this sketch. The photoelectric effect polarimeter (orange) uses 4 identical time projection chamber to track photoelectrons and presents a total of 31.2 cm of 0.25 atm dimethyl ether (DME) to the beam. The Compton polarimeter uses a 14 cm long scintillator (green) read out by a PMT (blue) to detect a Compton electron and the Compton scattered photon with one of 32 2mm thick CZT detectors (grey). The Compton polarimeter uses an active CsI shield (cyan) read out with a PMT (dark blue).

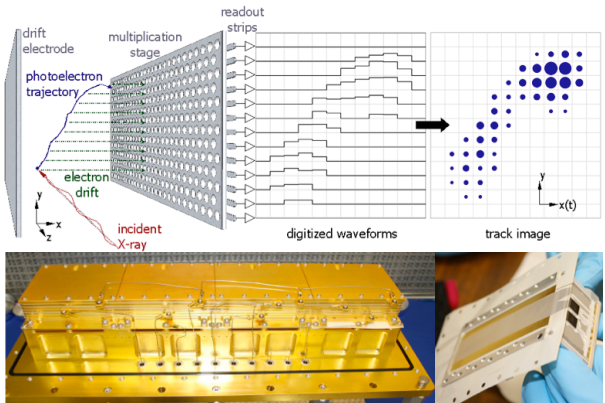

*Figure 6:* The photoeffect polarimeter uses gas electron multipliers to amplify the signals from the time projection chambers (upper image). The waveforms from several adjacent readout strips makes it possible to obtain a 2-D image of the photoelectron track. The lower images shows the prototype polarimeter for *GEMS* (left) and the readout strips (right)[1].

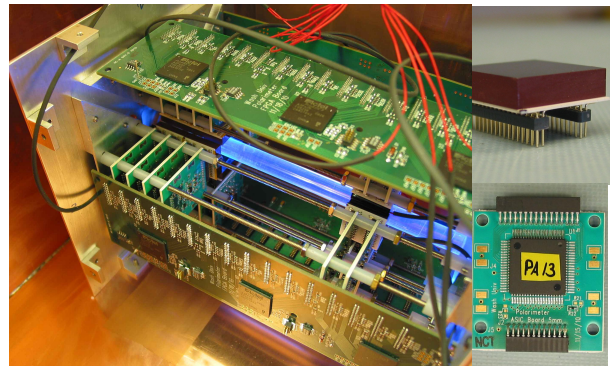

*Figure 7*: The left panel shows the *X-Calibur* polarimeter with the first 12 CZT detectors surrounding the scintillator rod. The other panels show CZT detectors bump bonded onto ceramic substrates with readout ASICs and readout electronics[3]. The *BEST* electronics will be smaller as one ASIC, rather than 2, will be used to read out each detector.

*3.4 Telescope and Mirror assembly:* Our design uses a deployable off-the-shelf optical bench to connect the spacecraft bus with the mirror assemblies similar to the one developed for *NuSTAR*. The mirror assemblies use the slumped glass mirror technology developed for *Con-X/IXO* and *NuSTAR* over the last decade (see Fig. 8). As of October 2011, the entire process of making, aligning, and bonding mirror segments has been developed and repeatedly demonstrated to achieve images better than 10″ half-power diameter at 4.5 keV. For *BEST*, the deposition of multi-layer coatings needs to be optimized to preserve the figure quality of the glass substrates. Sputtered thin films can produce significant stress (up to several giga-Pascal) and can create unacceptable figure distortion. We will use one of two methods to solve this problem: atomic layer deposition which creates opposite stresses on the two surfaces of the substrates canceling each other, or, appropriately select multi-layer materials with opposite stresses canceling each other. The alignment and bonding process that has been developed for *Con-X/IXO* is expected to work well for *BEST*, but needs to be verified for the relatively small shell spacing.

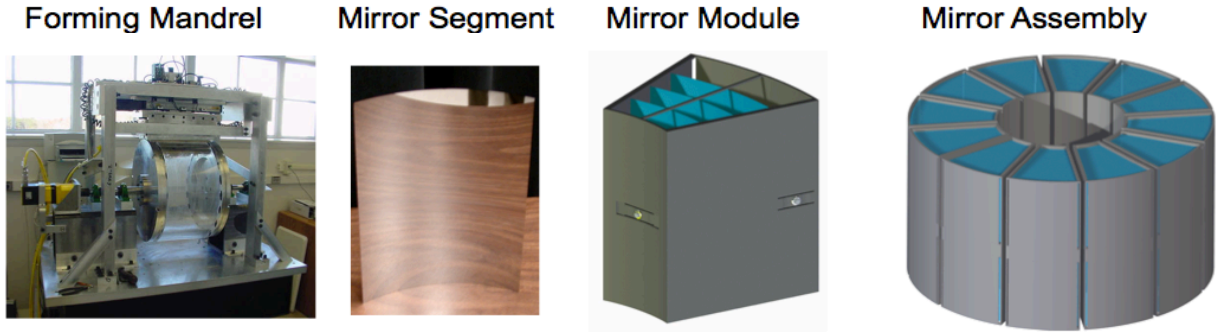

***Figure 8:*** The process of building the mirror assembly starts with forming mandrels whose precision optical figures are replicated to mirror segments. A number of mirror segments are precisely aligned and bonded together to become mirror modules, which are then aligned and integrated into a superstructure to become the flight mirror assembly.

**3.5 Spacecraft and Launch Vehicle:** *BEST* easily fits within the Taurus launch vehicle. We are conservatively baselining a 93″ fairing over a 63″ fairing because the maximum diameter of the observatory is 61″. With a mass estimate of 1040 kg, we will use the remaining Taurus launch capability of 1400 kg (600km orbit) to reduce the orbit inclination from the nominal 28° for reduced orbital background variations and minimized SAA passages. During launch the 10-m boom is stowed, and 1-2 weeks after launch the boom is deployed. The *BEST* mission and spacecraft specifications are given in Table 2. We plan for partial redundancy. The bus (patterned of the highly successful *Swift* bus design) will have redundant C&DH (2), Star Trackers (2), 6 reaction wheels, 2-axis gyros (3), and Power control. Unlike Swift, the wheels will be much smaller, because the high slew speeds are not required. The instruments are intrinsically redundant because each of the 6 polarimeters and 6 imagers is a stand-alone instrument with independent data, command, and power pathways from the bus.

***Table 2:*** *Spacecraft Parameters.*

| Imager | Requirement (Goal) |
|---|---|
| Mission life | 3 (5) years |
| Orbit altitude | 600 km, ≤28° |
| Radiation dose | <4 krad |
| Power bus/instrument | 223/358 W |
| Launch dimension | 1.6×2.2 m (Ø×height) |
| Telemetry (X-band) | 2.3 GB/day |
| Pointing control/stability/knowledge | 10″/3″/1″ (5″/2″/1″) |
| Slew speed | 0.2°/s |
| Settling time | 150 (130) s |
| Slews per week | 7-20 (depends on targets) |
| ToO response | 48 (36) hr |
| Field of regard | Sun angle 90°±20° |
| SAA operations? | No science (PMTs off) |

***3.6 Technical Readiness Level and Complexity:*** The main moveable components are the deployable optical bench, and the turntable to switch between imagers and polarimeters. The optical bench for *NuSTAR* is presently at TRL-5. A short development program to design and test a multiple mast configuration aimed at *BEST* will be able to quickly bring the level to TRL-6. The mirror technology will be at TRL-5 for the soft X-ray regime as soon as the long-term (~months) stability of the alignment of epoxy-bonded mirrors is verified in laboratory and vacuum environments. Elevating the technology for hard X-ray multilayer mirrors (presently at TRL-3) to the same TRL-level will require 1-2 years of targeted R&D. The technology of the hard X-ray imager is presently at TRL-3, that of the low-energy polarimeters at TRL-6, and that of the high-energy polarimeters at TRL-4. The *X-Calibur* balloon flight in 2014 will bring the high-energy polarimeter to TRL-6. *BEST* is of modest complexity. Each imager uses between 292 (requirement) and 584 (goal) ASIC channels. Each broadband polarimeter makes use of 4614 ASIC channels.

## *4 Operations*

***Orbital:*** On-orbit missions consist of stare-mode pointing (spinning about the bore sight) at a single target for a duration ranging from ½ to 1 day. There will be no change of targets during an orbit (to eliminate the dead-time from the Earth occultation). The high moment of inertia of the boom design preclude this kind of multi-target per orbit observing plan. Pointing can be true single-RA, Dec pointing, or a series of small delta RA-Dec tiling observations (i.e. ~10 arcmin between tiles). Additionally, the polarimeter detectors will have the PMT high-voltage turned off during SAA passages. Earth occultation plus the SAA results in a 45% live-time. ***Ground:*** The Mission Operations Center (MOC) will be at GSFC. For a ~28° inclination orbit, there will be 6-12 downlink possibilities per day (potential ground stations include USN-Hawaii/-Aus, GN-Santiago/-Wallops). Uplinks will be 1-2 times per week (for normal operations and the relatively rare Target of Opportunity (ToO) observations) given the long observing segments per target (an observing plan will cover 1 week, but can be uploaded more often for medium response ToOs). Rapid-response ToO can be uploaded any time via TDRSS. Level 1 processing of the telemetry will happen at the MOC, and Level 2 and 3 at the Science Operations Center (SOC).

## *5. Cost Estimate*

Cost for the *BEST* mission was done by using the total missions costs for the *Swift*, *NuSTAR*, and *GEMS* missions and the *Lobster* proposal and (a) subtracting delta costs for those for subsystems not on the *BEST* mission, (b) adding delta costs for new subsystems on *BEST* that are not on each of the comparison missions, and (c) scaling costs that are on both that change only in mass (e.g. bus structure mass); where "subsystems" means both instrument subsystems (i.e. different or new detectors) and bus-subsystems (e.g. redundancies needed to go from single-string Class C missions like *NuSTAR* and *GEMS* to redundant strings for a Class B mission like *BEST*). The results of those deltas and scalings are shown in Table 3. Using the 4 missions and proposal studies, the total mission costs for *BEST* range from 542 to 634 M$ (with an average of 582 M$). This cost includes the Taurus rocket. Since these cost estimates are based on existing missions, nearly existing missions and high-fidelity proposals, this cost range already includes reserves appropriate for this class and size mission (i.e. of order 30%) plus it includes the "wraps" for project management, systems engineering, quality assurance, flight software, I&T, and GSE/spares.

*Table 3*: Costing by comparison, scaling, adjusting from 4 missions & a proposal. Costs are in FY12 M$ and masses in kg. The columns "Delta Cost" and "Delta Mass" give the costs and masses, respectively, of BEST components not included in the costs of the comparison mission. The columns "Adjusted Cost" and "Adjusted Mass" give the total cost and mass estimates for BEST, respectively.

| Mission | Cost (then) | Year (then) | Cost (FY12) | Delta Cost | Adjust Cost | Mass [kg] | Delta Mass | Adjust Mass | Comments |
|---|---|---|---|---|---|---|---|---|---|
| *Swift* | 170 | 2004 | 167 | 335 | 592 | 1200 | 25 | 1225 | Existing |
| *Lobster* | 200 | 2010 | 206 | 368 | 634 | 398 | 520 | 988 | Proposal |
| *NuSTAR* | 110 | 2009 | 117 | 365 | 542 | 360 | 602 | 1022 | L-4 mo |
| *GEMS* | 125 | 2011 | 125 | 370 | 555 | 313 | 552 | 925 | In Phase B |
| **BEST** | n/a | n/a | n/a | n/a | **582** | n/a | n/a | **1040** | **Average** |

We estimate the uncertainty on these scaling calculations is +/- 25%. De-scoping from 6 to 5 (4) telescopes would reduce the cost by ~45 M$ (90 M$) and can be used to keep the mission safely in the 300-600M$ cost bracket.

## 6. Management Plan / Schedule / Decommissioning:

*BEST* would be a University PI-lead mission (H. Krawczynski, Wash. Univ.) with GSFC responsible for the project management, systems engineering, instrument and systems I&T, and mission operations. Observatory-level I&T can be done either at the s/c bus vendor facilities or at GSFC. Mirror design and fabrication will be done at GSFC (W. Zhang, lead). The polarimeter detectors will be designed and built at Washington U. and GSFC, and the imaging detector will designed and built at GSFC (J.Tueller, lead). Science operations will be done at Washington Univ. The mission program will take 5 years from beginning of Phase A to the end of Phase D with the following breakdown: 9, 15, 18, and 18 months, respectively. Phase E is 3 yrs (with a 5-yr goal). Decommissioning will take 3 months by natural orbital decay. The reentry products will be less than 7 m$^2$.